# Design and Fabrication of an Optimum Peripheral Region for Low Gain Avalanche Detectors

Pablo Fernández-Martínez, D. Flores, S. Hidalgo, V. Greco, A. Merlos, G. Pellegrini and D. Quirion

Instituto de Microelectrónica de Barcelona, IMB-CNM (CSIC), Campus UAB, 08193, Cerdanyola del Vallès, Barcelona, Spain

Corresponding autor: D. Flores, david.flores@imb-cnm.csic.es

**Abstract**

Low Gain Avalanche Detectors (LGAD) represent a remarkable advance in high energy particle detection, since they provide a moderate increase (gain ~10) of the collected charge, thus leading to a notable improvement of the signal-to-noise ratio, which largely extends the possible application of Silicon detectors beyond their present working field. The optimum detection performance requires a careful implementation of the multiplication junction, in order to obtain the desired gain on the read out signal, but also a proper design of the edge termination and the peripheral region, which prevents the LGAD detectors from premature breakdown and large leakage current.

This work deals with the critical technological aspects when optimising the LGAD structure. The impact of several design strategies for the device periphery is evaluated with the aid of TCAD simulations, and compared with the experimental results obtained from the first LGAD prototypes fabricated at the IMB-CNM clean room. Solutions for the peripheral region improvement are also provided.

## 1. Introduction

The Low Gain Avalanche Detector (LGAD) structure [1], plotted in Fig. 1, is based on the standard PiN diode architecture. The generated charge multiplication of LGADs is provided by adding a moderately doped P-type diffusion (P-Well) beneath the highly doped N-type electrode, which increases the doping concentration in the vicinity of the $N^+P$ junction with respect to the highly resistive substrate ($\rho$ = 10 k$\Omega$·cm). Due to P-Well diffusion, the electric field at the junction experiences a notable increase under reversed bias conditions, to such an extent that the impact ionization mechanism allows electrons generated by the incident radiation to undergo avalanche multiplication before being collected. In this sense, the LGAD performance is analogous to that of the Avalanche Photo-Diode [2] (APD), regularly used for optical and X-ray detection [3, 4]. However, LGAD detectors aims at lower gain values on the output signal (in the range of 10, against typically > 100, for APD), which makes them more suitable for the tracking detection of high-energy charged particles. In the LGAD detectors the initial current is only moderately amplified without significant increase of the noise level, thus improving the signal-to-noise (S/N) ratio. High gain detectors produce a large amount of charge due to multiplication that cannot be efficiently collected by the readout electronics with trigger times in the range of 25 ns. In addition, LGAD detectors offer the possibility of having fine segmentation pitches, thus allowing the fabrication of micro-stripped or pixelated devices, which do not suffer from readout cross-talk between adjacent elements. Segmented LGAD devices will only have edge termination at the last cell since the inner ones are self protected.

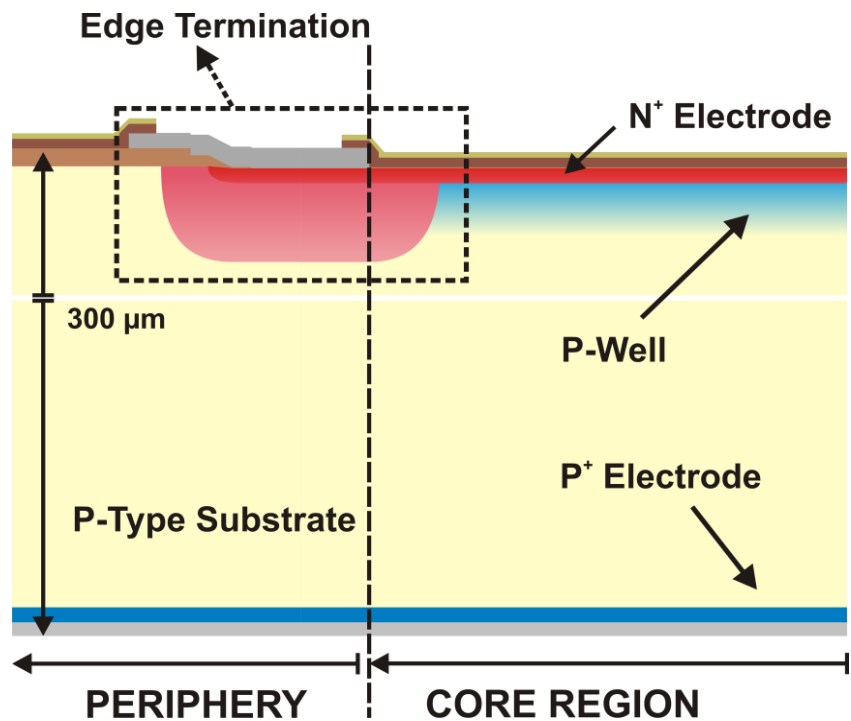

**Fig. 1.** Cross-section of the LGAD structure (half-cell): The $P^+/\pi/P/N^+$ core region, where the multiplication takes place, covers the central area of the device. The less efficient peripheral region includes the edge termination to prevent a premature breakdown and structures to minimize the surface leakage current.

As depicted in Fig. 1, the $P^+/\pi/P/N^+$ junction resulting from the typical LGAD structure is only implemented in the central region of the device (core region), in such a way that the area covered by the P-Well diffusion can be associated with the active area for detection. The peripheral region, extending from the core to the device edge, is considered a less efficient area in terms of detection, although it plays a crucial role to ensure the stability and uniformity of the electric field distribution within the core region. Besides, the peripheral region contributes to the reduction of the undesired leakage current collected in the N-type electrode of the detector that degrades the signal-to-noise ratio. In this sense, the design of the periphery has to provide a maximum effectiveness in terms of surface leakage reduction, optimum electric field at the core and a breakdown voltage in excess of 1000 V, which is the limit of the read-out electronics in typical applications.

In addition, the multiplication junction requires an optimised edge termination strategy in order to prevent a premature breakdown, which would spoil the use of LGAD detectors in many applications. The performance of different edge termination designs is analysed in Section 2, by evaluating the electric field profile with the TCAD simulation software [5]. In the same way, Section 3 is dedicated to the effect of positive oxide charges and the possible technical solutions to be implemented in the peripheral region to eliminate the surface inversion layer and the subsequent high leakage current levels. Experimental data of fabricated LGAD detectors is provided in Section 4, including the capacitive behaviour. Finally, the main conclusions of the work are summarised in Section 5. The work reported in this paper has been performed in the framework of the CERN RD50 collaboration.

## 2. Edge termination of the multiplication junction

The multiplication junction of the LGAD detector is created by an initial Boron implantation and the subsequent high temperature anneal followed by a high dose Phosphorous or Arsenic implantation with a low temperature anneal. As a consequence, a deep P-well diffusion with a peak concentration in the range of $1 \times 10^{16}$ cm$^{-3}$ and a shallow $N^+$ electrode diffusion are formed, leading to a $P^+/\pi/P/N^+$ structure. The resulting doping profile makes possible that, under reverse bias conditions, the electric field at the $N^+P$ junction rises up to a value high enough to activate the impact ionization mechanism, which leads to charge multiplication However, as the same mechanism leads to the avalanche that can eventually cause the junction breakdown, the presence of the P-Well results in a significant reduction of the detector capability to stand high voltages with respect to a conventional PiN [6] design implemented on the same substrate with identical process technology except the P-Well Boron implantation.

The electric field increase is particularly critical at the $N^+P$ junction edges, where the junction shows a cylindrical curvature, as a consequence of the planar fabrication process [7]. This behavior can be observed in Fig. 2, where the simulated electric field and electrostatic potential distributions at the $N^+P$ junction edge are shown for the case of an LGAD detector biased at a typical operational reverse voltage of 400 V. The electrostatic potential, crowding at the junction curvature, lead to a local increase of the electric field in this region.

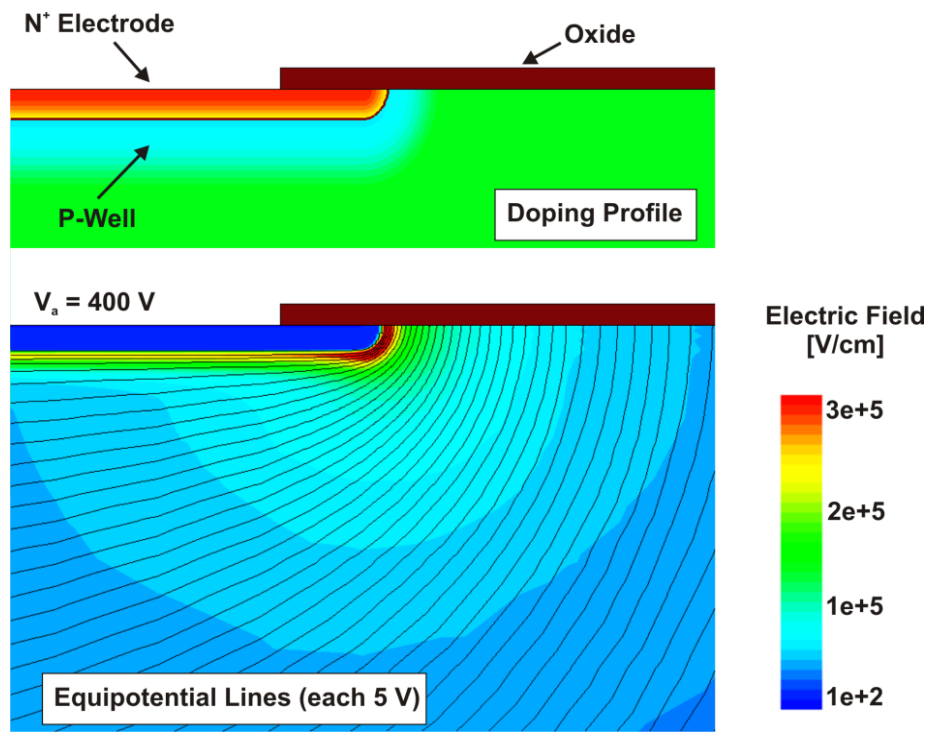

**Fig. 2.** Simulated doping profile (top) and electric field combined with equipotential lines distribution (bottom) at the edge termination of an unprotected LGAD multiplication junction at a reverse bias of 400 V.

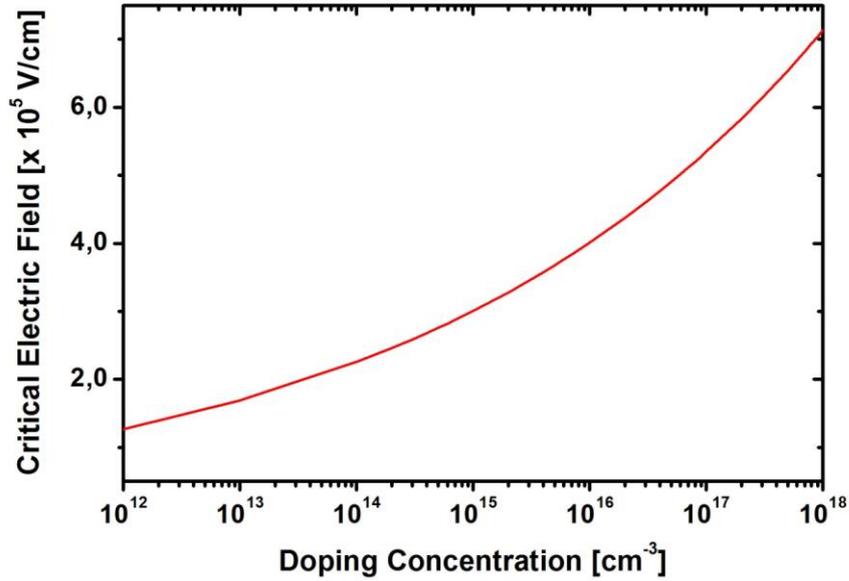

**Fig. 3.** Critical electric field as a function of the doping concentration, according to Baliga [7].

The use of extremely high resistivity substrates to ensure full depletion of the detector at relatively low voltage has a direct impact on the critical electric field ($E_C$) value at which avalanche takes place in the different regions of the LGAD structure. The lower doped side of the $N^+P$ junction has a peak concentration in the range of $1\times10^{16}$ cm$^{-3}$. Taking into account the dependence of the critical electric field on the doping concentration plotted in Fig. 3, derived from [7], the $E_C$ value is in the range of $4\times10^5$ V/cm. However, if no P-well diffusion is present, as in standard PiN detectors, the $N^+P^-$ junction, with a lower doping concentration in the range of $1\times10^{12}$ cm$^{-3}$, has an $E_C$ value reduced to $1.5\times10^5$ V/cm. Therefore, the optimization of the core region edge has to consider these different values and take benefit from them to balance the breakdown point.

The modification of the electric field distribution at the junction edges might compromise the uniformity of the multiplication, as well. Charge carriers collected through the edge can have different multiplication value than those collected through the core region, where the electric field is uniform. In finely-segmented devices, in which the edge and core areas become comparable, this problem can lead to a large non-uniformity in the signal collection. In fact, uniformity is usually the main goal when optimizing a detector design. In this sense, the breakdown voltage of LGAD detectors ($V_{BD}$) has to be high enough to ensure the full depletion ($V_{FD}$) of the substrate ($V_{BD} > V_{FD}$) and the activation of the multiplication mechanisms. As a matter of fact, $V_{BD} = 1000$ V offers a safety margin to operate 300 µm-thick pad LGAD detectors; even if they are intended to work in harsh radiation environments, for which $V_{FD}$ significantly increases as a consequence of the radiation damage [8].

Taking this into account, three main requirements are identified when designing an optimum edge termination for the multiplication junction:

1) The maximum voltage sustained by the planar junction should not be limited by the edge termination. Thus:

$$V_{BD|edge} > V_{BD|planar} \qquad (1)$$

where, $V_{BD|edge}$ and $V_{BD|planar}$ are the breakdown voltage values for the edge termination and the planar region, respectively.

2) The electric field distribution has to be uniform all over the multiplication junction, confining the impact ionization process to the core region.

3) The process technology must be compatible with a standard production of medium or small area (< 1 cm²) pad detectors, as well as with the segmentation of the electrodes, in order to obtain strip or pixel LGAD detectors.

Among the various available solutions, three different designs have been analyzed to meet the previous requirements: a floating guard ring surrounding the multiplication junction, the extension of the $N^+$ shallow diffusion beyond the mask limits of the P-Well diffusion, and the use of an N-type deep diffusion overlapping the junction edge.

*A) Floating Guard Ring*

This design uses a floating guard ring placed at a given distance ($W_R$) from the junction edge, being its width in the range of 30 μm. The N-type floating ring is implemented using the $N^+$ electrode shallow diffusion, with no additional photolithographic process. When the main junction is reverse biased, the voltage along the ring is fixed at a certain value and the electric field is then distributed to a larger area with less crowding at the ring curvature. Although the implanted dose is equal, the junction depth of the ring becomes higher, as a consequence of enhanced Phosphorous diffusion into the extremely low doped P-type substrate [9]. Fig. 4 shows the simulated final doping distribution, together with the electric field and electrostatic potential distributions when the reverse bias is set to 400 V. As soon as the depleted region reaches the N-type floating ring, it becomes biased by punch-through [7] at an intermediate voltage value, causing a redistribution of the electrostatic potential in the edge termination and the appearance of a second electric field peak at the ring curvature.

The resulting electric field distribution can be observed in Fig. 5, where the horizontal section of the electric field along the junction edge is compared with that obtained for an unprotected junction biased at the same reversed voltage ($V_a = 400$ V). The double peak distribution in the guard ring approach leads to a significant reduction of the electric field at the curvature of the main junction. Depending on the $W_R$ value, both peaks can be modulated to achieve almost the same value, enlarging the voltage sustained by the termination.

However, as the N-type ring is directly diffused on the substrate, which doping concentration is several orders of magnitude lower than that of the P-Well diffusion, a significantly lower $E_C$ value is expected at the ring edge, according to the $E_C$ doping dependence described in Fig. 3. In this sense, the requirement for an optimum edge termination, described in Eq. (1), can hardy be satisfied, since a premature avalanche breakdown will start first at the ring curvature. Besides, the resulting electric field distribution is not uniform and the guard ring strategy is not advisable for segmented designs, as it would increase insensitive area of the detector.

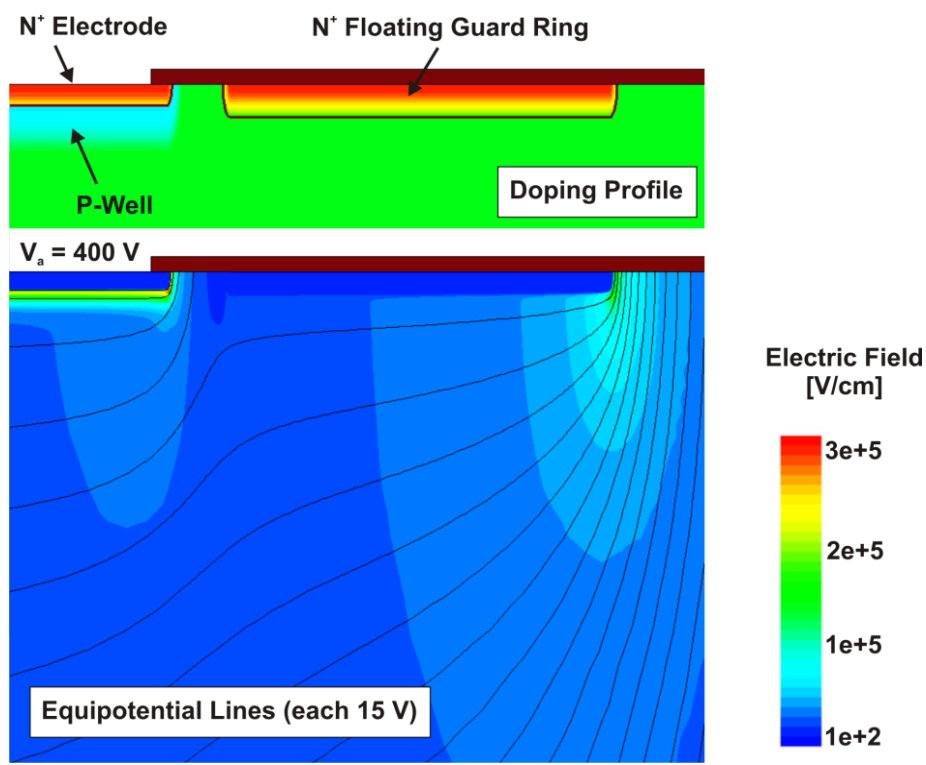

**Fig. 4.** Simulated doping profile (top) and electric field combined with equipotential lines distribution (bottom) at a reverse bias of 400 V when a shallow N-type floating ring is implemented.

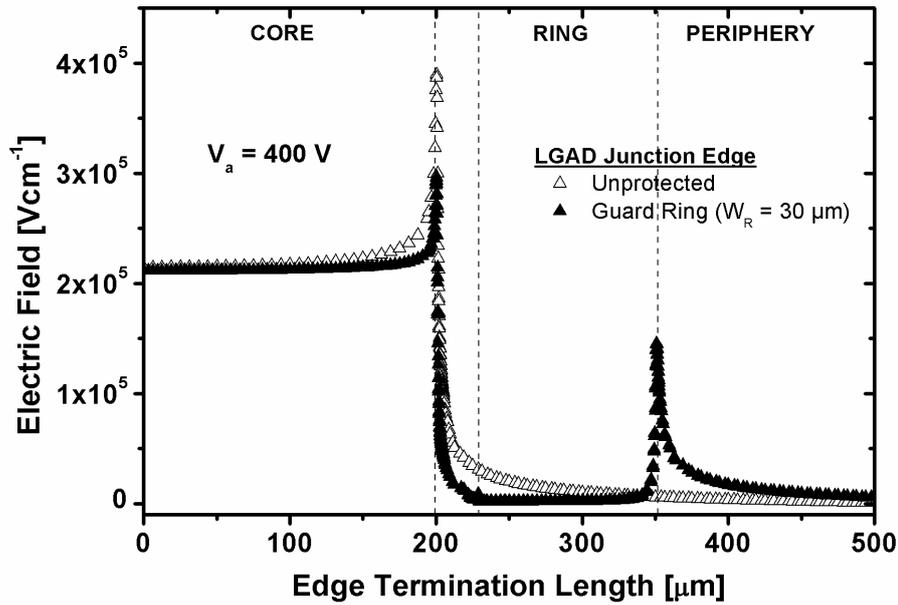

**Fig. 5.** Simulated electric field profile at the shallow $N^+$ junction depth for an unprotected multiplication junction and for the same junction protected with a floating guard ring at a reverse bias of 400 V.

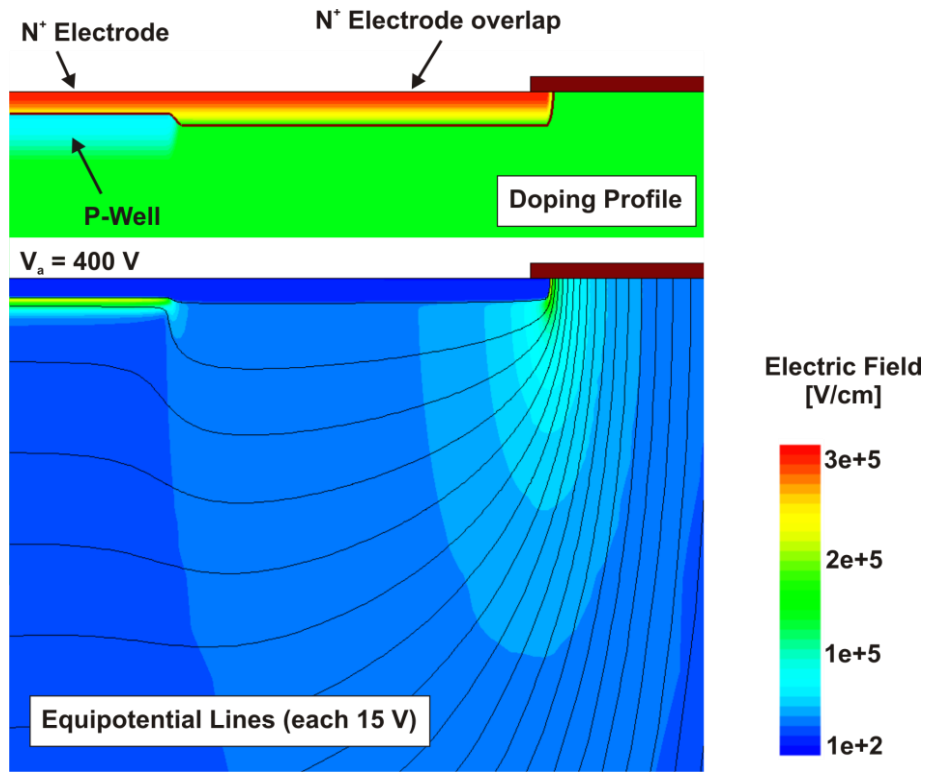

**Fig. 6.** Simulated doping profile (top) and electric field combined with equipotential lines distribution (bottom) at a reverse bias of 400 V when the $N^+$ electrode extension is implemented.

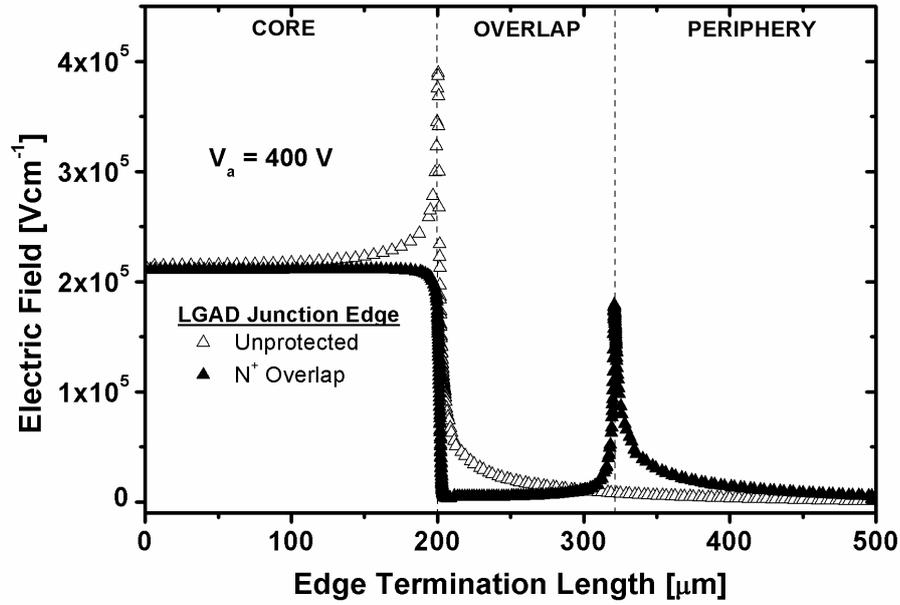

**Fig. 7.** Simulated electric field profile at the shallow N$^+$ junction depth for an edge termination consisting of an extension of the N$^+$ electrode diffusion beyond the junction curvature at a reverse bias of 400 V.

*B) N$^+$ Electrode Extension*

The second edge termination approach is implemented by extending the shallow N$^+$ electrode diffusion beyond the mask limits of the P-Well, in such a way that it overlaps the original curvature of the P/N$^+$ junction. As depicted in Fig. 6, diffusion through the lowly doped substrate results in a deeper N$^+$ junction in the overlapping region, with the subsequent reduced curvature. The simulated electric field and electrostatic potential distributions, also plotted in Fig. 6 under a reverse bias of 400 V, clearly show how the equipotential lines are not anymore crowded at the multiplication junction edge, making possible a uniform increase of the electric field at the planar junction of the core region. The electric field profile at the shallow N$^+$ junction depth at $V_a$ = 400 V, reported in Fig. 7, shows the electric field peak at the N$^+$ extension edge and identical electric field value in the planar junction as in the case of the floating ring approach. Comparing Figs. 5 and 7, it is clear that the use of a floating guard ring does not remove the electric field peak at the edge of the planar junction (Fig. 5). On the contrary, this electric field peak is completely removed when the N$^+$ extension is implemented, according to Fig. 7. Nevertheless, the electric field peak at the edge of the N$^+$ extension is much higher than that of the planar junction when the reverse bias approaches the breakdown value, which is in the range of 800 V. As a consequence, the maximum impact ionization is not located at the planar region. Finally, the cross-section of the fabricated edge termination with the shallow N$^+$ diffusion extension is provided in Fig. 8. Assuming that the $E_C$ value is lower in this region, this peak makes again difficult to fulfil the first requirement for an optimum edge termination structure. In conclusion, the extension of the shallow $N^+$ diffusion does not provide a high breakdown voltage although it is compatible with strip or pixel designs.

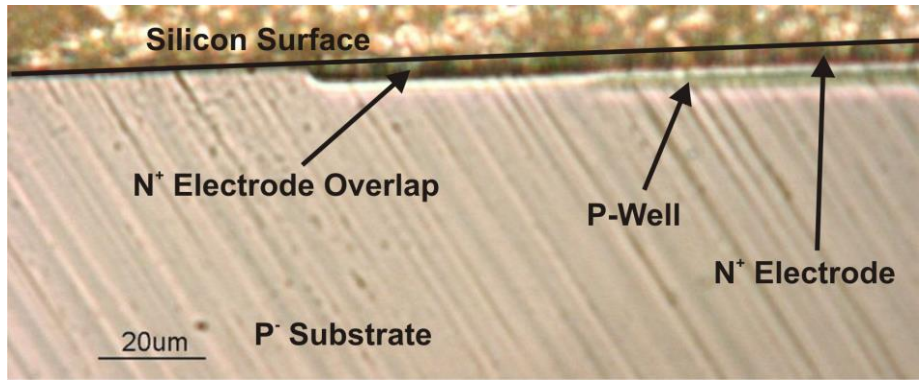

**Fig. 8.** Cross-section of the edge termination region when the extension of the $N^+$ electrode is considered.

*C) Deep N-type diffusion*

This termination approach is based on the inclusion of a deep N-type diffusion at the edge of the multiplication region connected to the $N^+$ electrode diffusion, as shown in Fig. 9. The deep N-type diffusion requires an additional mask level and the implantation dose has to be carefully tuned to allow the depleted region to partially spread into the new diffusion, as in the well-known Junction Termination Extension (JTE) [10] technique widely used in power devices. The effectiveness of the deep N-type diffusion can be enhanced when combined with a metal field plate, provided the field oxide has the required thickness. The electric field and equipotential distributions plotted in Fig. 9 corroborate the reduction of the electric field at the deep N-type diffusion curvature. Indeed, the electric field profile at the $N^+$ junction depth reported in Fig. 10 at $V_a = 400$ V, where the $N^+$ extension is compared with the deep N-type diffusion with and without field plate, clearly shows that the electric field peak at the curvature of the deep N-type diffusion is lower than that of the planar junction. This condition is maintained when the device is driven to breakdown (>1000 V) and the maximum ionization is always placed at the planar junction, as desired. Now the $E_C$ value at the edge termination is not as low as in the previous cases since the doping difference between the deep N-type diffusion and the substrate is not as high as in the $N^+$ extension case [7]. Therefore, the initial requirement of the optimum edge termination is fulfilled. The cross-section of the fabricated edge termination with a deep N-type diffusion is provided in Fig. 11.

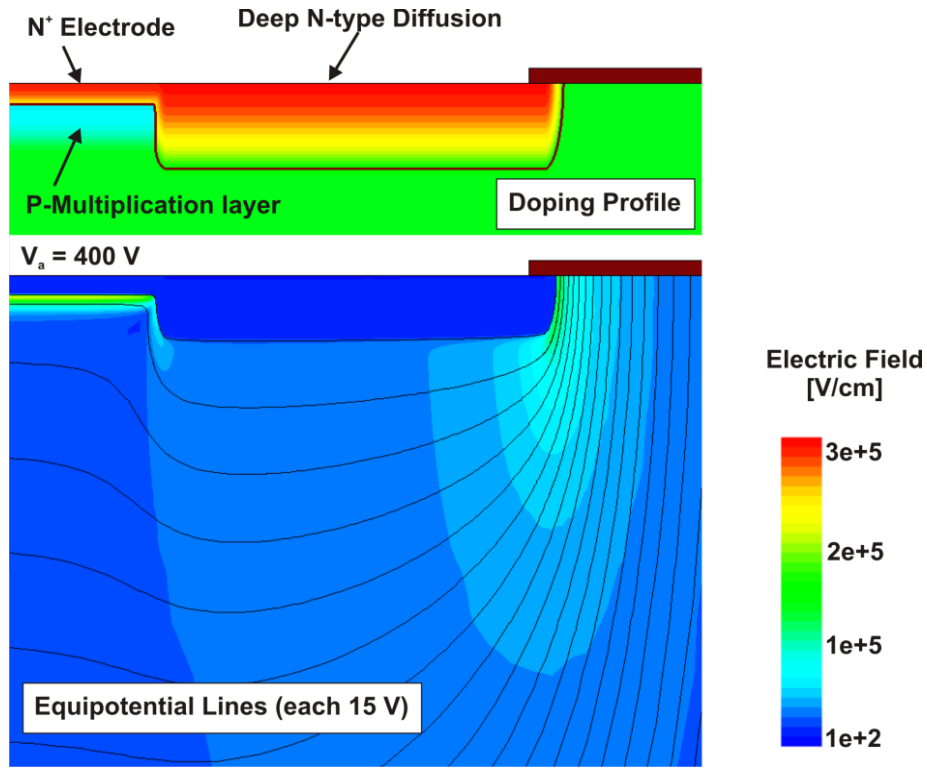

**Fig. 9.** Simulated doping profile (top) and electric field combined with equipotential lines distribution (bottom) at a reverse bias of 400 V when the deep N-type diffusion is implemented.

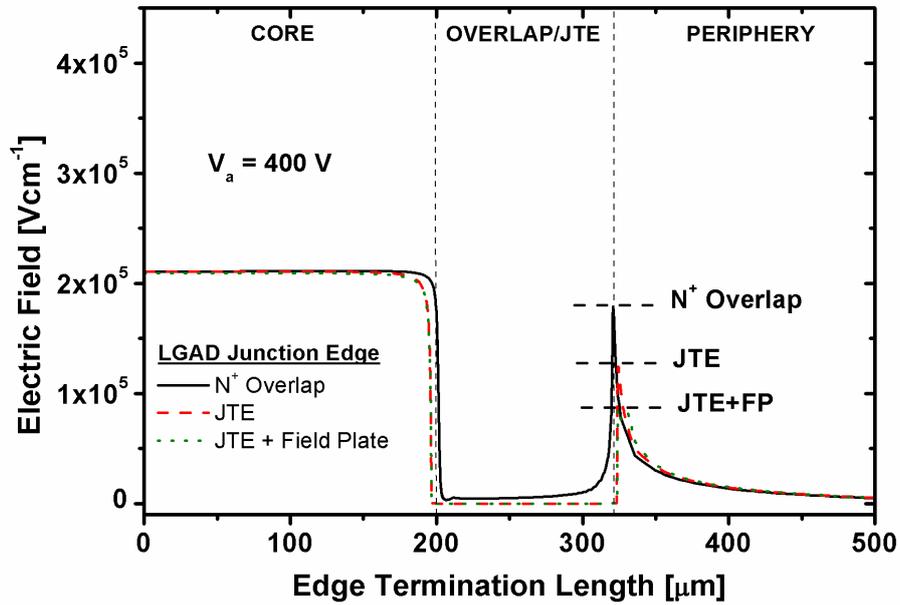

**Fig. 10.** Simulated electric field profile at the shallow $N^+$ junction depth for a multiplication junction protected by the extension of the $N^+$ electrode ($N^+$ overlap) and by a deep N-type diffusion (JTE) with and without Field Plate (FP) at a reverse bias of 400 V.

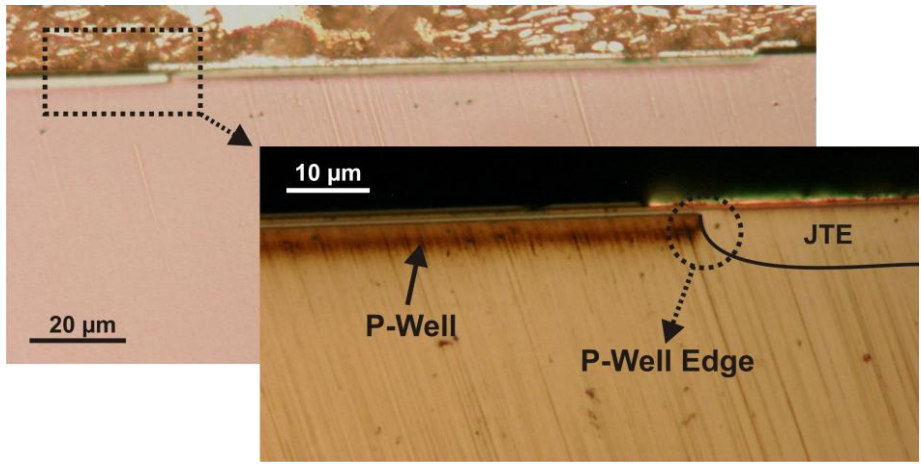

**Fig. 11.** Cross-section of the edge termination region with a deep N-type diffusion.

## 3. Peripheral region

The design of the peripheral region is crucial to obtain high performing LGAD detectors able to sustain high voltage values with minimum leakage current injection into the core region. The optimization of the peripheral region has to be carefully done, according to the following aspects.

### A) General aspects

First, it is convenient to provide the peripheral region with enough Silicon area to prevent the depletion region from reaching the dice edge. Typically damaged by the dicing process, the lateral chip surfaces contain a high concentration of structural defects, which constitute an important source of leakage current. Under increasing reverse bias, the LGAD substrate is depleted both in the vertical and lateral directions. As it is exemplified in Fig. 12, the lateral depletion of the substrate becomes relevant after the full depletion in the vertical direction at the $V_{FD}$ value. Once this value is reached, the additional supplied voltage is then supported by the lateral spread of the depletion region. If the peripheral region is too short, the depletion region can eventually reach the chip edge, leading to a high current injection into the core region. A common design rule to avoid this undesirable effect lies in extending the peripheral length at least twice the thickness of the substrate (i.e. > 600 μm, for the actual 300 μm LGAD substrates).

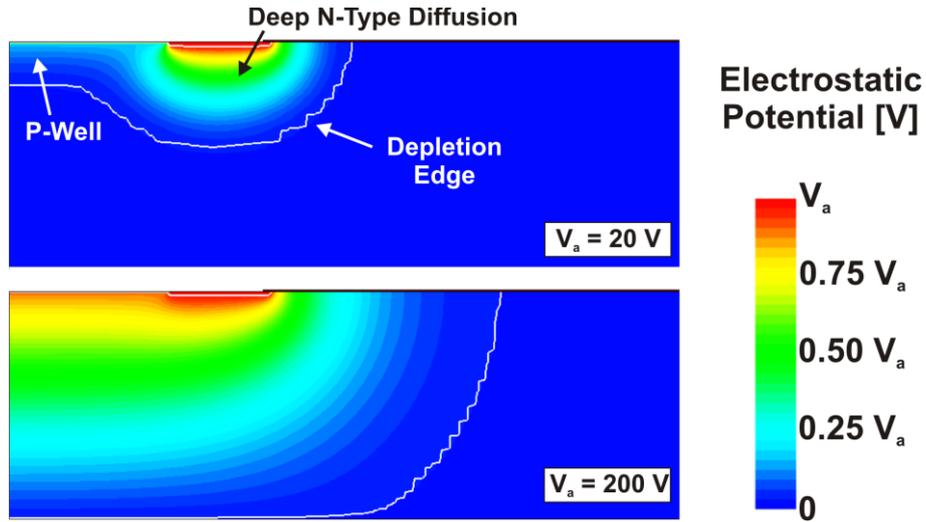

**Fig. 12.** Simulated depletion dynamics in the peripheral region of an LGAD detector at reverse bias of 20 V ($< V_{FD}$) and 200 V ($> V_{FD}$).

Nevertheless, even with a properly dimensioned periphery, transient overvoltage events may lead the depletion region to reach the chip edge, thus compromising the detector stability. Although these events hardly happen in detector applications, they are very common in power electronics, where the subsequent high current levels can cause the device thermal destruction. Power devices are usually protected with a so-called channel stopper [11]: a highly doped diffusion, with the same metallurgical character as the substrate, used to stop the lateral spread of the depletion region at the Silicon surface. Placed at the LGAD chip edge, the channel stopper also prevents the build-up of parasitic current paths along the peripheral surface, typical when high resistive P-type substrates (10 kΩ·cm, in this case) are used and an inversion channel is created at the Silicon surface due to the inherent positive oxide charges.

*B) Positive oxide charges*

The periphery is usually covered with a thermal $SiO_2$ layer, the thickness of which is determined by the metal field plate structure. During the detector fabrication, the so-called field oxide is used as an implantation mask to protect the LGAD surface from undesired doping penetration. After the fabrication, the field oxide becomes a passivation layer preventing the Silicon substrate from contamination and electrostatic discharges. Although the oxide growth process includes different techniques to minimize the contaminant ions, the field oxide typically contains a certain concentration of positive fixed charges within its volume [12], which induce the accumulation of negative charges along the surface of the peripheral region and the subsequent inversion layer of the very low doped P-type substrate ($10^{12}$ cm$^{-3}$).

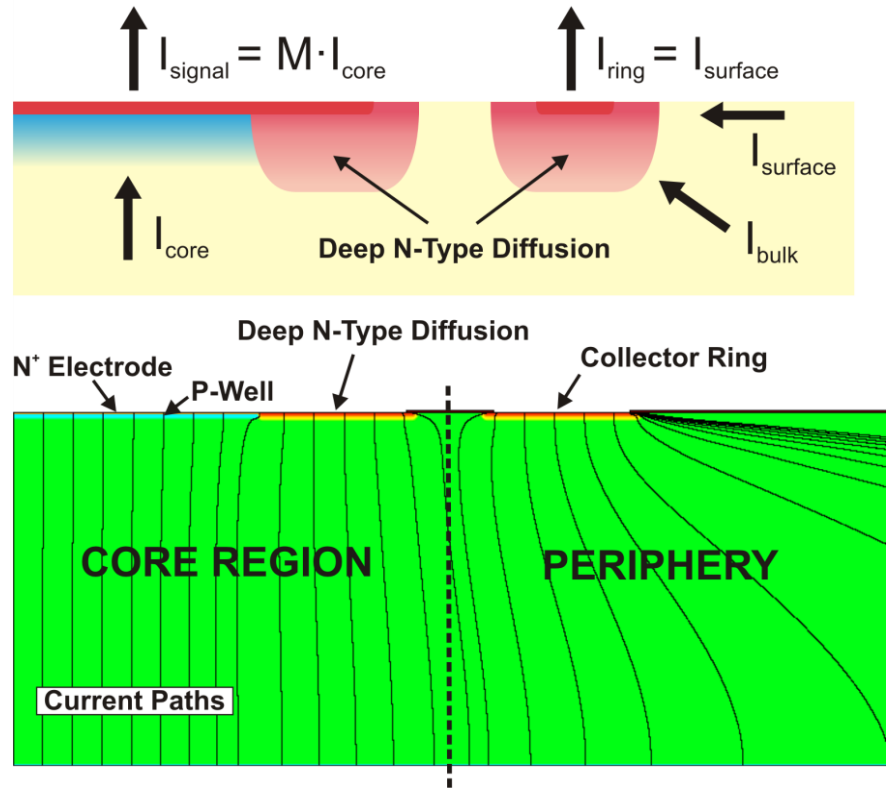

**Fig. 13.** Schematic operation of an LGAD detector provided with a collector ring (top) and simulated current paths (bottom).

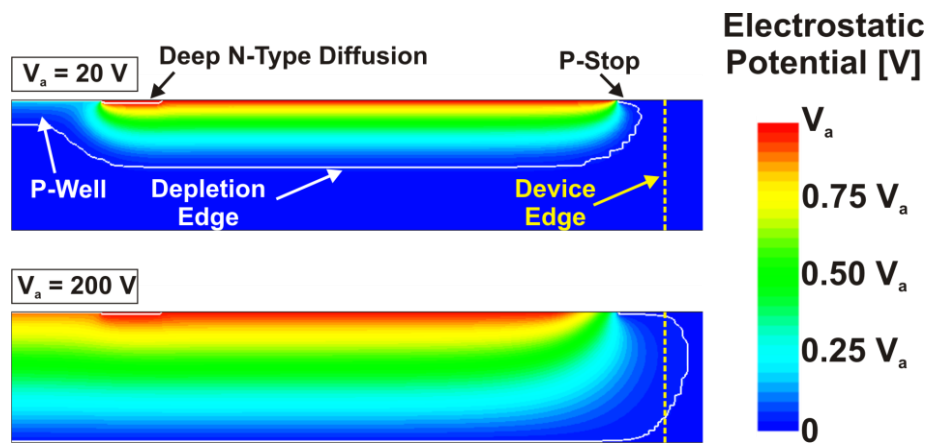

**Fig. 14.** Depletion dynamics with a 1e11 cm$^{-2}$ of positive fixed charges in the field oxide.

The build-up of an N-type channel all through the substrate surface is a major challenge when fabricating detectors on P-type Silicon, since the different implemented N-type structures (strips, pixels, pad electrodes or rings) may result electrically connected. As a matter of fact, in pad LGAD detectors the collector ring structure used to separately collect the leakage current generated at the periphery of the device, described in Fig. 13, becomes useless, and some of the edge termination strategies (i.e. floating guard ring) lose their effectiveness. In addition, the depletion dynamics within the peripheral region is severely modified. Fig. 14 shows the simulated extension of the space charge region in the LGAD periphery when a positive charge density of 1e11 cm$^{-2}$ within the field oxide volume is considered. For a reverse bias well below the V$_{FD}$ value (V$_a$ = 20 V), the whole peripheral surface is at the same electrostatic potential, since the N$^+$ electrode is now virtually extended through the inversion channel up to the channel stopper. At a higher reverse bias (V$_a$ = 200 V), the depleted region reaches the chip edge, regardless of the peripheral region length.

The virtual extension of the N-type electrode, if no protection techniques are implemented on the Silicon surface, modifies the LGAD charge collection dynamics, since the whole peripheral volume becomes integrated into the active detection area. Carriers generated in the peripheral region do not undergo multiplication since they are not collected through the multiplication junction. Hence, the detector readout may show two superposed signals: one with multiplication, corresponding to the carriers collected through the multiplication junction in the core region, and a second one, without multiplication, corresponding to the carriers generated in the peripheral region. In a typical LGAD design with an effective detection area of 25 mm² and a total device area of 64 mm², the contribution to the signal from the peripheral region can be as high as the multiplied current since the dicing of the chips produces a large number of defects which generate leakage current levels much higher than those of bulk Silicon.

The virtual extension of the N-type electrode has also a relevant effect on the device capacitance. In full depletion conditions, the capacitance can be expressed as:

$$C = \varepsilon \cdot A/d \quad (2)$$

where d is the thickness of the depleted substrate and A is the total area where the depleted region is spread. If no charge is considered within the field oxide, A is basically the N$^+$ electrode area, with a minor increase due to the lateral advance of the depleted region. Hence, C is almost constant for V$_a$ > V$_{FD}$.

*C) Collector ring*

The surface leakage current in high resistive substrates significantly degrades the signal-to-noise ratio in radiation detectors. Therefore, many detector designs include an additional electrode for a differentiate extraction of the surface leakage current. In LGAD designs, this additional electrode is implemented as a biased N-type ring, placed in the peripheral region, as shown in Fig. 13. The biased ring — designated as collector ring in this work, to avoid misunderstanding with the floating guard ring discussed in the previous section — is implemented with the same deep N-type diffusion used in the edge termination structure and placed close to the core region edge to avoid the sustaining voltage capability degradation. To preserve the edge termination efficiency, identical voltage values have to be applied to both the $N^+$ electrode and the collector ring, whereas the read out is performed independently. The simulated current paths depicted in Fig. 13 corroborate that the collector ring collects the bulk and the surface leakage currents generated within the detector periphery. Thus, only the current generated within the core region is collected by the $N^+$ electrode. The collector ring also plays a relevant role in the capacitance performance of the LGAD detector. The capacitance of the core region when the collector ring is used decreases up to the full depletion value ($C[V_{FD}] = C_{FD}$). Then, the lateral spread of the depletion region and its impact on the capacitance is measured as the capacitance of the collector ring, facilitating the match between the detector and the readout electronics.

*D) P-Spray and P-Stop*

The inclusion of the collector ring has to be accompanied with technical solutions to avoid an electrical connection between the detection electrode and the added ring through to the surface inversion layer due to the positive fixed charges in the field oxide. In this sense, two solutions have been considered in the design of the peripheral region: the P-Spray [13] and the P-Stop [14], as drawn in Fig. 15. The P-Spray is based on an initial blanket Boron implantation with low dose and low energy, to create a shallow low doped P-type layer at the surface (P-Spray) thus avoiding the creation of the inversion layer. However, this solution requires a precise control of the implantation process and the total thermal budget. The first processed wafers with P-Spray (no collector ring included) exhibited abnormal capacitance behavior due to the non-controllable doping profile of this shallow P-type layer. The P-Stop is created by using the channel stopper implantation and is the most effective way to eliminate the surface current path at the cost of added peripheral area.

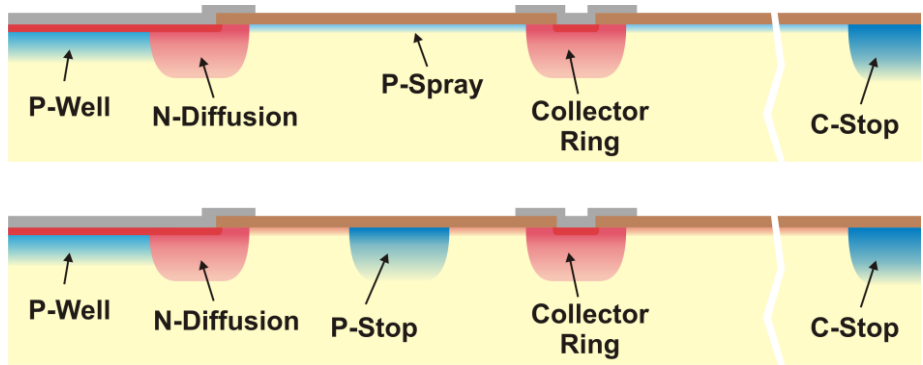

**Fig. 15.** Technical solutions to avoid the surface inversion effect in very low doped P-type substrates due to the positive fixed charges in the field oxide.

## 4. Experimental Results

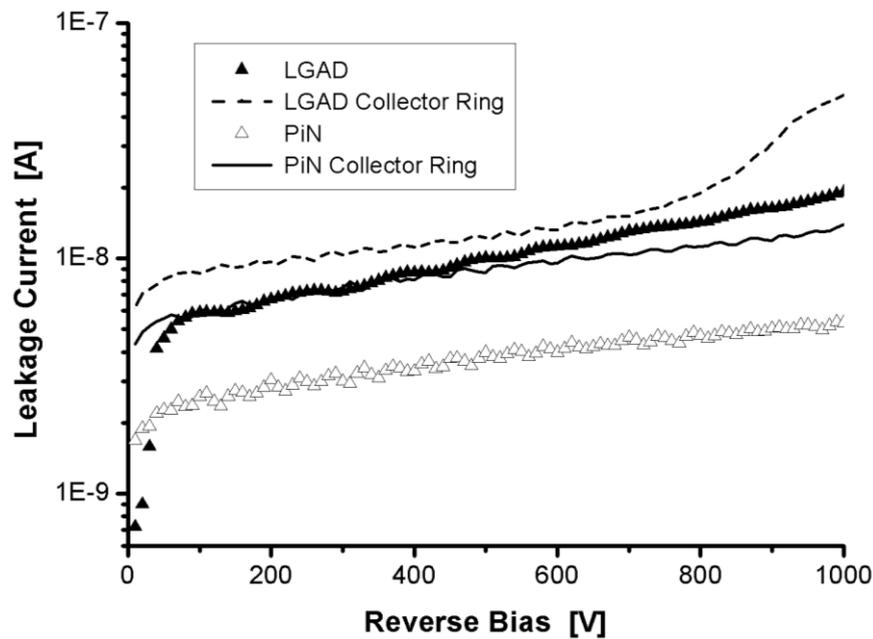

**Fig. 16.** Experimental I(V) characteristics of equivalent LGAD and PiN detectors.

LGAD detectors have been fabricated with a gain in the range of 10 and a capability of sustaining more than 1000 V with different detection areas [1]. The edge termination and peripheral regions were optimized according to the analysis reported in this paper. Therefore, a deep N-type diffusion is used to protect the multiplication region and to implement the collector ring. The P-stop diffusion is also included to eliminate the electrical connection between N-type diffusions due to the surface inversion layer. The leakage current of the detection region and the collector ring current up to 1000 V are reported in Fig. 16 where it can be inferred that the LGAD detector exhibits a 5 times higher leakage current than that of the corresponding PiN diode counterpart at a reverse bias of 400 V due to the inherent multiplication effect. This ratio cannot be directly seen as the detector gain which is always defined by the collected charge ratio during the transient event.

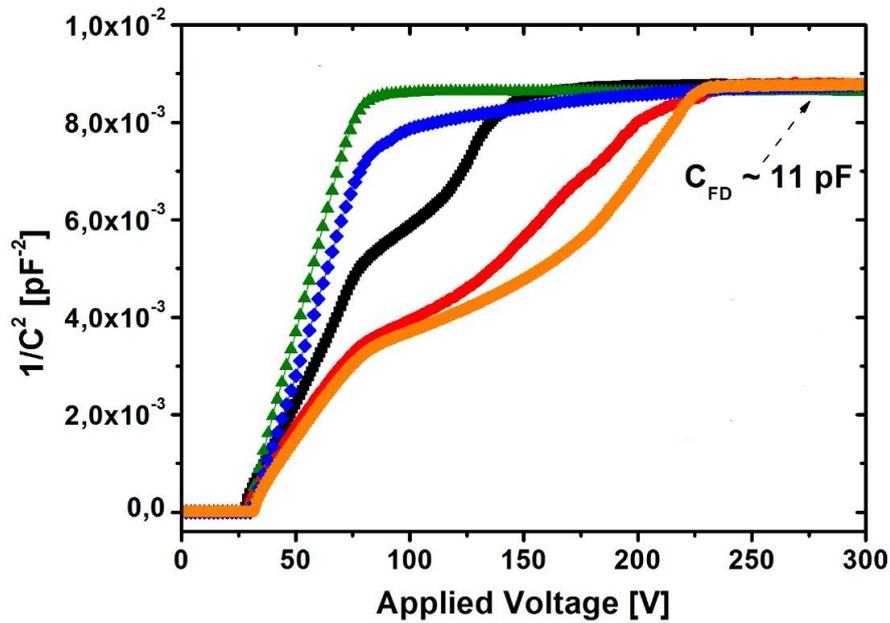

**Fig. 17.** Experimental capacitance performance (I/C²) of fabricated LGAD detectors with P-Spray to avoid surface inversion.

The evolution of 1/C² versus the applied reverse bias is plotted in Fig. 17 for different LGAD samples with different oxide charges. Measures were performed at 20ºC with an amplitude of 500 mV and a frequency of 10 kHz (according to CERN standards). An initial high capacitance value is found until the P-type multiplication layer is completely depleted. Then, the capacitance decreases with the applied reverse bias until the full depletion of the substrate. From $V_{FD}$ on, the capacitance is almost constant ($C_{FD}$ = 11 pF) with a little contribution of the lateral spread of the depletion region. It is worth to mention that the ideal capacitance of the detection region is in the range of 9 pF. The bumps in some of the 1/C² curves are due to the inefficient P-Spray diffusion, not enough doped to avoid surface inversion when the P-Stop is not present.

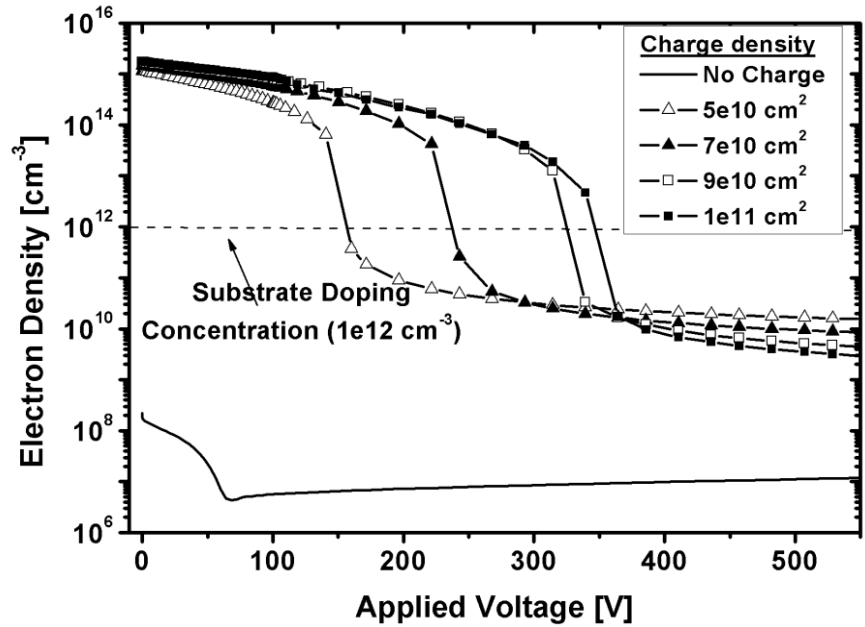

**Fig. 18.** Simulated electron density at the Silicon surface of the peripheral region.

The simulated electron density close to the Silicon surface of the peripheral region as a function of the positive oxide charges and the applied reverse bias is reported in Fig. 18, where no P-Spray diffusion is implemented. On the contrary, positive oxide charges create a shallow inversion layer which becomes finally depleted at a high reverse bias, increasing with the oxide charge density. Other technological solutions to minimize the oxide charge effect can be studied and optimized. The use of a different dielectric on top of the peripheral region with reduced or even negative charge density could enhance the detector performance without using any additional P-type diffusion at the peripheral region.

## 5. Conclusions

The optimum electrical performances of Silicon LGAD detectors require, a part from an accurate control of the P-type multiplication layer, a careful optimization of the edge termination and the peripheral region to be sure that no premature breakdown or high surface leakage current levels will arise. Three approaches are contemplated for the edge termination structure, with the deep N-type diffusion as the best performing solution in terms of electric field profile at the Silicon surface. The design of the peripheral region is driven by the effect of the inherent positive oxide charges which create a surface inversion layer in the extremely low doped P-type substrates. Therefore, two technological solutions have been tested (P-Spray and P-Stop) to eliminate the surface current path. Experimental results (leakage current and capacitance) corroborate the voltage sustaining capability in excess of 1000 V and the effectiveness of the implemented edge termination and peripheral region.

**Acknowledgments**